\documentclass[]{aastex701}
\usepackage{url}

\begin{document}

\title{Twenty-five years of XMM-Newton EPIC-MOS2 Unfocused Particle Background: Solar Cycle Modulation, Galactic Cosmic-Ray Correlation, and Implications for NewAthena}

\author[orcid=0000-0003-0852-0257,sname='Vaia']{B. Vaia}
\affiliation{Istituto Nazionale di Astrofisica, Istituto di Astrofisica Spaziale e Fisica Cosmica di Milano, via A. Corti 12, 20133 Milano, Italy}
\email[show]{beatrice.vaia@inaf.it}  

\author[orcid=0000-0002-9112-0184,sname='Gastaldello']{F. Gastaldello}
\affiliation{Istituto Nazionale di Astrofisica, Istituto di Astrofisica Spaziale e Fisica Cosmica di Milano, via A. Corti 12, 20133 Milano, Italy}
\email{fabio.gastaldello@inaf.it}

\author[orcid=0000-0002-2483-278X,sname='Molendi']{S. Molendi}
\affiliation{Istituto Nazionale di Astrofisica, Istituto di Astrofisica Spaziale e Fisica Cosmica di Milano, via A. Corti 12, 20133 Milano, Italy}
\email{silvano.molendi@inaf.it}

\author[orcid=0000-0001-7703-9040,sname='Bartalucci']{I. Bartalucci}
\affiliation{Dipartimento di Fisica, Università degli studi di Roma Tor Vergata, Via della Ricerca Scientifica 1, I-00133 Roma, Italy}
\affiliation{Istituto Nazionale di Astrofisica, Istituto di Astrofisica Spaziale e Fisica Cosmica di Milano, via A. Corti 12, 20133 Milano, Italy}
\email{iacopo.bartalucci@inaf.it}

\author[orcid=0009-0001-3048-0020,sname='Balboni']{M. Balboni}
\affiliation{DIFA - Universit\`a di Bologna, via Gobetti 93/2, I-40129 Bologna, Italy}
\affiliation{Istituto Nazionale di Astrofisica, Istituto di Astrofisica Spaziale e Fisica Cosmica di Milano, via A. Corti 12, 20133 Milano, Italy}
\email{marco.balboni@inaf.it}

\author[orcid=0000-0003-0879-7328,sname='Ghizzardi']{S. Ghizzardi}
\affiliation{Istituto Nazionale di Astrofisica, Istituto di Astrofisica Spaziale e Fisica Cosmica di Milano, via A. Corti 12, 20133 Milano, Italy}
\email{simona.ghizzardi@inaf.it}

\author[orcid=0000-0002-3754-2415,sname='Lovisari']{L. Lovisari}
\affiliation{Istituto Nazionale di Astrofisica, Istituto di Astrofisica Spaziale e Fisica Cosmica di Milano, via A. Corti 12, 20133 Milano, Italy}
\email{lorenzo.lovisari@inaf.it}

\author[orcid=0000-0002-8017-0338,sname='Marelli']{M. Marelli}
\affiliation{Istituto Nazionale di Astrofisica, Istituto di Astrofisica Spaziale e Fisica Cosmica di Milano, via A. Corti 12, 20133 Milano, Italy}
\email{martino.marelli@inaf.it}

\author[orcid=0000-0002-9775-732X,sname='Rossetti']{M. Rossetti}
\affiliation{Istituto Nazionale di Astrofisica, Istituto di Astrofisica Spaziale e Fisica Cosmica di Milano, via A. Corti 12, 20133 Milano, Italy}
\email{mariachiara.rossetti@inaf.it}

\author[0000-0002-6038-1090]{A. Tiengo}
\affiliation{Scuola Universitaria Superiore IUSS Pavia, Piazza della Vittoria 15, 27100 Pavia, Italy}
\affiliation{Istituto Nazionale di Astrofisica, Istituto di Astrofisica Spaziale e Fisica Cosmica di Milano, via A. Corti 12, 20133 Milano, Italy}
\email{andrea.tiengo@iusspavia.it}

\author[0000-0002-7887-1485]{C. Macculi}
\affiliation{INAF, Istituto di Astrofisica e Planetologia Spaziali (IAPS), Via del Fosso del Cavaliere 100, 00133 Roma, Italy}
\email{claudio.macculi@inaf.it}

\author[0000-0002-3088-1561]{S. Lotti}
\affiliation{INAF, Istituto di Astrofisica e Planetologia Spaziali (IAPS), Via del Fosso del Cavaliere 100, 00133 Roma, Italy}
\email{simone.lotti@inaf.it}

\author[0000-0002-3088-1561]{V. Fioretti}
\affiliation{INAF, Osservatorio di Astrofisica e Scienza dello Spazio (OAS) di Bologna, Via P. Gobetti 93/3, 40129 Bologna, Italy}
\email{valentina.fioretti@inaf.it}

\author[0000-0002-6082-5384]{T. Mineo}
\affiliation{INAF, Istituto di Astrofisica Spaziale e Fisica Cosmica (IASF) di Palermo, Via Ugo La Malfa 153, 90146 Palermo, Italy}
\email{teresa.mineo@inaf.it}

\author[0000-0002-4962-0740]{G. Schellenberger}
\affiliation{Center for Astrophysics, Harvard \& Smithsonian, 60 Garden Street, Cambridge, MA 02138, USA}
\email{gerrit.schellenberger@cfa.harvard.edu}

\author[0000-0002-4737-1373]{C. E. Grant}
\affiliation{Kavli Institute for Astrophysics and Space Research, Massachusetts Institute of Technology, 77 Massachusetts Avenue, Cambridge, MA 02139, USA}
\email{cgrant@mit.edu}

\begin{abstract}

We present a study of the unfocused particle-induced background measured by the XMM–Newton EPIC-MOS2 detector, extending previous analyses to the end of 2025 and covering approximately twenty-five years of observations. Using out-of-field-of-view (outFOV) data, we investigate the long-term evolution of the particle background over two complete solar cycles and compare it with independent tracers of the Galactic cosmic-rays. We find an extremely tight correlation between the MOS2 outFOV count rate and the AMS proton flux, with an intrinsic scatter of only about 1.1\%, confirming that variations in the unfocused background are primarily driven by changes in the Galactic cosmic-rays flux. The extended temporal baseline clearly reveals the modulation associated with the solar cycle and the characteristic signatures of the 22 year Hale cycle. A Lomb-Scargle analysis identifies a dominant periodicity at 11.28 yr, while structure-function and wavelet analyses reveal a persistent $\sim$27 day modulation associated with solar rotation. We also confirm the presence of a residual background component that does not scale with the Galactic cosmic-ray flux, consistent with previous studies and possibly associated with Compton-scattered high-energy X-ray photons. This work places the particle-induced background measured by the XMM–Newton EPIC-MOS2 detector in the broader context of Galactic cosmic-ray studies, while also providing direct implications for the development of background-monitoring and mitigation strategies for future X-ray missions such as \textit{NewAthena}.

\end{abstract}



\section{Introduction}\label{sec1}
\setcounter{footnote}{0}

The success of next-generation missions, such as the European Space Agency's Advanced \textit{NewAthena}\footnote{\url{https://www.the-athena-x-ray-observatory.eu/en}}, depend in no small manner on our ability to characterize the instrumental background, as it can dominate the signal from faint, diffuse emission. The official ESA requirements for \textit{NewAthena} call for the particle-induced background to be known to a precision of 2\%{}\footnote{see page 43, third row of \url{https://www.cosmos.esa.int/documents/400752/507693/Athena_SciRd_iss1v5.pdf}}, highlighting the need for an accurate understanding of its temporal variability.
Scheduled for launch in 2039, \textit{NewAthena} will provide unprecedented sensitivity and spectral capabilities for the study of the hot and energetic Universe \citep{Natureathena}.
The \textit{NewAthena} scientific payload consists of a 12-m focal length X-ray telescope equipped with two instruments: the X-ray Integral Field Unit (X-IFU; \citealt{Barret2018, Peille2025}), designed for high spectral-resolution imaging, and the Wide Field Imager (WFI; \citealt{Meidinger2018}), which will provide wide-field imaging and mid-resolution spectroscopy in the 0.2--15 keV energy range.

The direct predecessor of \textit{NewAthena} within the ESA science programme is the European Space Agency's X-ray Multi-Mirror satellite \textit{XMM-Newton} \citep{citeXMM} launched in 1999 in High Earth Orbit (HEO).
The \textit{XMM-Newton} satellite carries three European Photon Imaging Cameras (EPIC, \citealt{MOScamera}; \citealt{PNcamera}) that operates in the energy range between 0.2 and 12 keV.
Since the WFI aboard \textit{NewAthena} will operate in a similar energy range with silicon-based detectors, the two instruments will share similar instrumental effects. This makes the study of the \textit{XMM-Newton} background particularly relevant for \textit{NewAthena}, providing moreover valuable insight into both the particle environment and its temporal variability important for the characterization and mitigation of the background component for both mission instruments (X-IFU and WFI).


This work focuses on the particle-induced background, namely the background generated when high-energy Galactic Cosmic Rays (GCRs, mostly protons), with energies above approximately 100 MeV, interact with the spacecraft and detector materials.
The signals from the primary cosmic-ray particles themselves are usually removed by the onboard electronics because they deposit very large amounts of energy or create characteristic charge patterns across multiple pixels.
However, these interactions also produce secondary particles -- mainly electrons and photons -- that leave signals in the detector similar to genuine astrophysical X-ray events. These secondary signals are difficult to filter out and therefore contribute to the detector background \citep{Hall2008,vonKienlin2018}.

The GCR flux -- and therefore the background component produced by it -- strongly anticorrelate with the 11-year solar cycle, peaking during solar minimum, when the heliospheric magnetic field provides reduced shielding. However, because the polarity of the large-scale solar magnetic field reverses every solar cycle, the complete modulation pattern follows the 22-year Hale cycle, leading to observable differences between consecutive solar cycles. 

The study of the unfocused component of the background has been carried out in the past years through several different approaches. These include the analysis of detector regions not exposed to the sky (outFOV) and Filter Wheel Closed (FWC) observations (e.g. \citealt{GastaldelloSP}), which provide direct measurements of the particle-induced background free from both astrophysical X-ray photons and soft protons which are instead concentrated by the telescope mirrors. In addition, onboard particle-rejection diagnostics, such as high-energy reject rates and the EPIC-pn number of discarded lines parameter, have been widely used as indirect tracers of the flux of charged particles interacting with the detectors and spacecraft structure (e.g. \citealt{Sarkar2024}). A further approach exploits EPIC-pn observations performed in Small Window Mode, where the onboard particle filtering is disabled, allowing the direct study of particle tracks produced by high-energy cosmic rays in the detector (e.g. \citealt{Schellenberger2025}).

In this work, we investigate the \textit{XMM-Newton} unfocused particle background over two complete solar cycles using the MOS2 outFOV signal.
The goal is to characterize the long-term evolution of the unfocused particle background in order to develop and optimize mitigation strategies for the WFI aboard \textit{NewAthena}.
This work was developed within the framework of the ESA R\&D project ATHENA Radiation Environment Models and X-ray Background Effects Simulators (AREMBES, Contract No.: 4000116655/16/NL/BW). Here, within the same framework, we extend the analysis performed by \citet{GastaldelloSP}, based on  the first 15 years of \textit{XMM-Newton} operations, up to the end of 2025 covering for the first time 25 years of \textit{XMM-Newton} observations. 

The paper is organized as follows. In Section \ref{sec2} we describe the analysis of the \textit{XMM-Newton} MOS2 outFOV data set. In Section \ref{sec3} we present the external datasets used for comparison with the \textit{XMM-Newton} measurements: the \textit{Chandra} ACIS reject-rate data and the AMS-02 proton flux measurements. In Section \ref{sec4} we quantitatively investigate the correlations between the different datasets, search for possible time lags in the different datasets, and study the presence of short-timescale periodicities in the MOS2 outFOV signal. Finally, the results are discussed in Section \ref{sec5} and summarized in Section \ref{sec6}.

\section{The \textit{XMM-Newton} data set}\label{sec2}

Following the approach adopted by \citet{GastaldelloSP}, MOS2 outFOV light curves are extracted from all available \textit{XMM-Newton} MOS observations collected between 1 January 2015 and 1 January 2026, 6718 exposures ($\sim 222$ Ms in 4744 revolutions).

The MOS cameras are generally preferred over the pn camera for studies of the unfocused particle background because the pn detector does not provide a truly outFOV region. The area conventionally referred to as outFOV is partially contaminated by focused X-ray photons and soft protons \citep{Marelli2021}. Moreover, MOS1 is not used in this work because two of its outer CCDs were damaged early in the mission. Therefore, in this analysis we use only data from the MOS2 camera.

\begin{figure}
    \centering
    \includegraphics[width=0.8\textwidth]{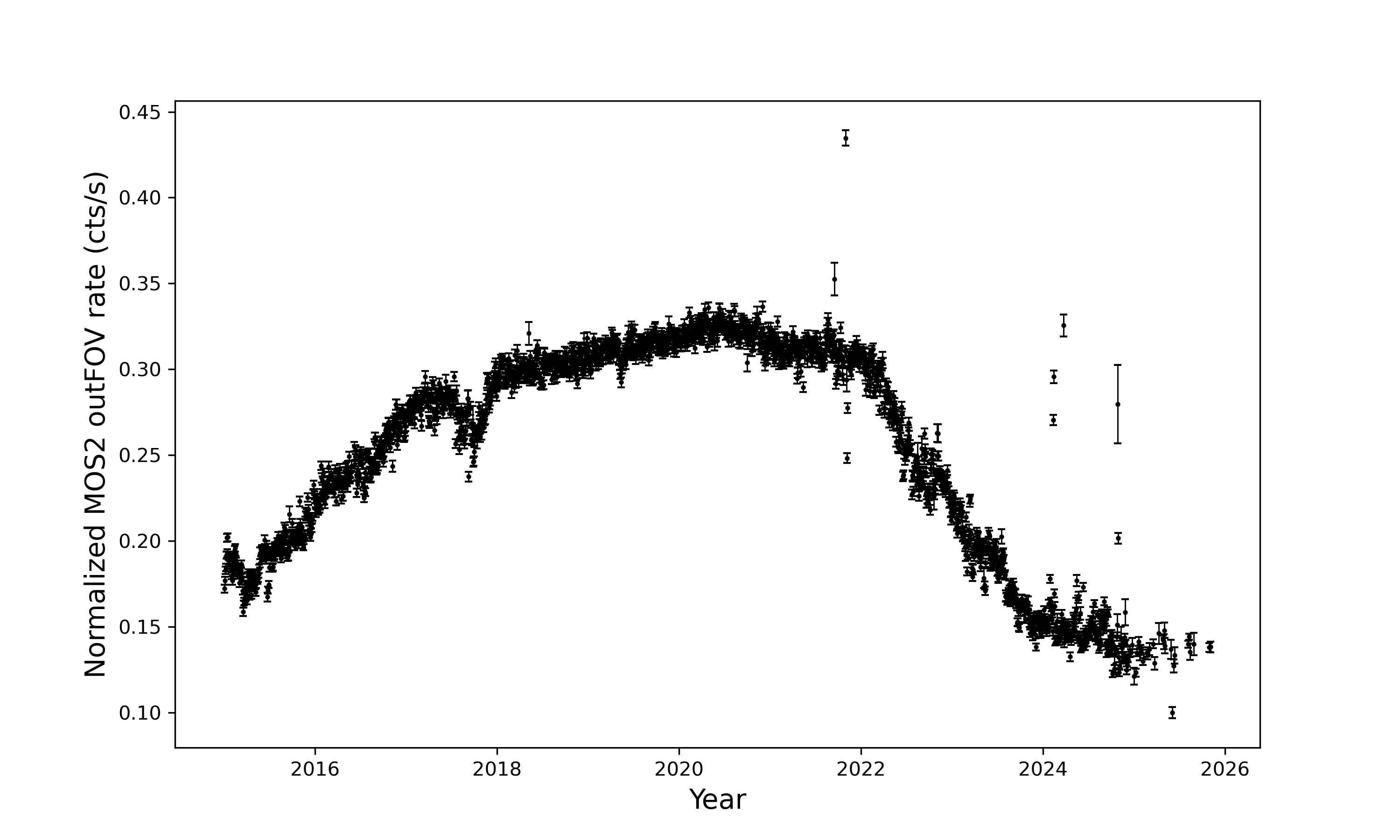}
    \caption{MOS2 outFOV count rate from January 2015 to December 2025, renormalized to the full field-of-view area following \citet{GastaldelloSP}, as a function of time. Each point corresponds to a single \textit{XMM-Newton} revolution.}
    \label{fig:lc_2015_2026}
\end{figure}

We applied the same event patterns, flag selections, energy filters, and outFOV extraction region adopted in \citet{Marelli2017} and \citet{GastaldelloSP}\footnote{\texttt{
(PATTERN<=12)\&\&
((PI in [7000:9400]) || (PI in [10000:11000]))\&\&
(FLAG==0)} and region selection as in Section 3 of \citet{Marelli2017}.}
For each observation, outFOV light curves are produced with time bins of 1000 s. Individual bins with fewer than 20 counts are removed in order to retain only intervals with reliable count rates and uncertainties. This selection removes approximately 3\% of the total exposure. The resulting light curves are then grouped according to the corresponding \textit{XMM-Newton} revolution ($\sim 48$ h). Fig. \ref{fig:lc_2015_2026} shows the full dataset after this binning procedure. 
In particular, the outFOV count rate is re-normalized to the geometrical area of the full field of view by multiplying the measured outFOV rate by the ratio between the inFOV and outFOV detector areas. 

Since the background tends to increase near the end of each revolution because of the spacecraft entrance into the radiation belts (see Fig. 3 in \citealt{GastaldelloSP}), we excluded the final portion of each orbit by removing data acquired at orbital phases larger than 0.8, following the approach adopted by \citet{GastaldelloSP}. The orbital phase was computed using the TSTART and TSTOP keywords of the orbit files available for each \textit{XMM-Newton} revolution on the \textit{XMM-Newton} webpage. These files are produced by the \textit{XMM-Newton} Science Analysis Software (SAS, version 20.0.0) orbit task with a sampling interval of 1 s.

Major Solar Energetic Particle (SEP; upward outliers of the general trend in Fig. \ref{fig:lc_2015_2026}) events are removed after having generated a SEP list from proton flux measured by GOES satellites \citep{SW_cite} at energies greater than 10 MeV. SEP intervals were identified whenever the proton flux exceeded the threshold adopted by the ESA Solar Energetic Particle Environment Modelling (SEPEM) application server, namely 0.001 $\rm{cm^{-2}}\,\rm{s^{-1}}\,\rm{sr^{-1}}\,\rm{MeV^{-1}}$. In this way, we excluded 98 revolutions contaminated by SEP events ($\sim 6\%$ of the total number of revolutions in our sample between 2015 and 2026).

In addition, revolutions still showing significant residual variability after the previous filtering steps are excluded from the final sample. Specifically, we reject revolutions for which the probability of the light curve being fitted by a constant is lower than $10^{-5}$. Only 2 revolutions are excluded in this phase.

\begin{figure*}
    \centering
    \includegraphics[width=0.8\textwidth]{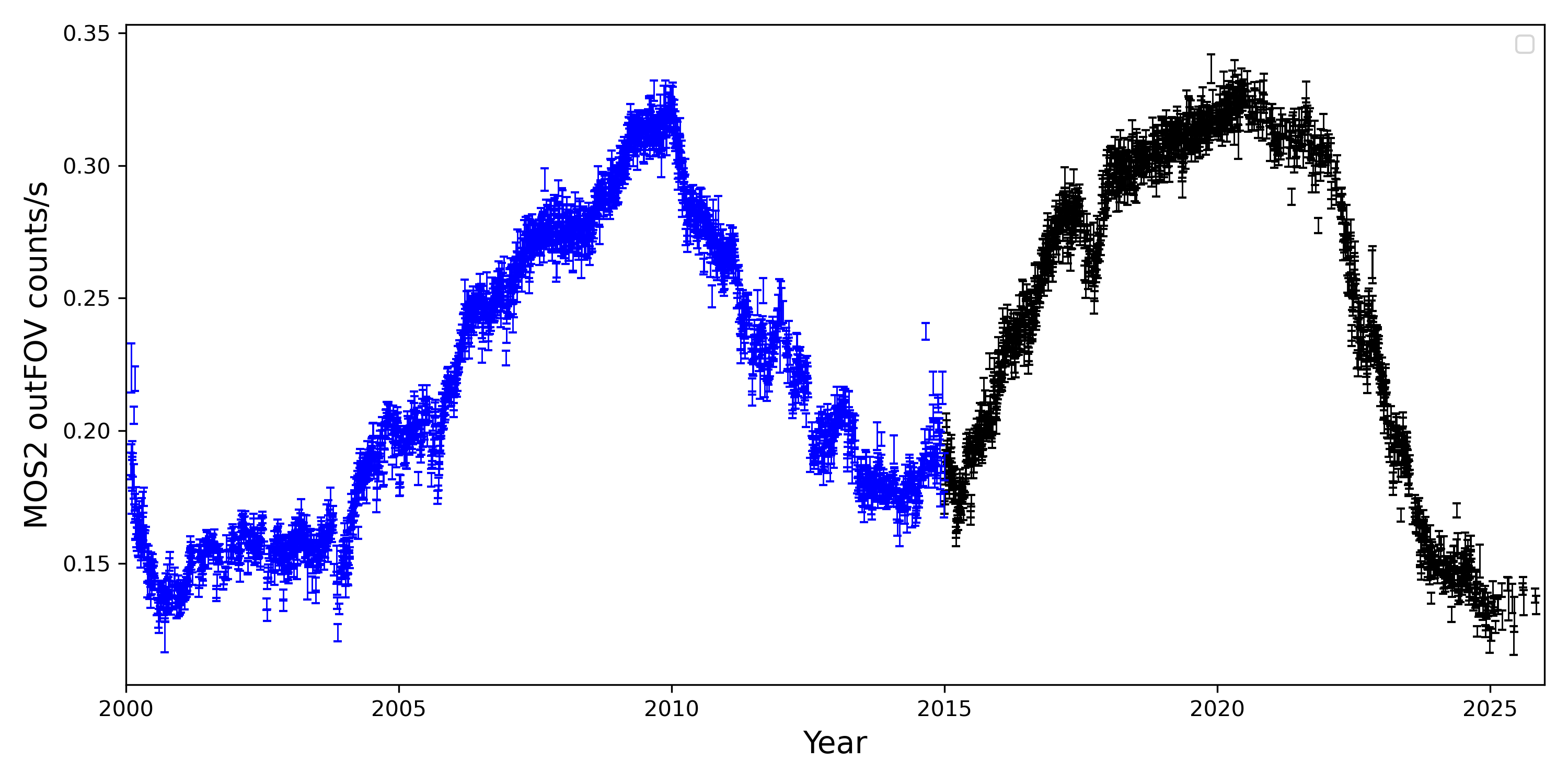}
    \caption{MOS2 outFOV count rate from January 2000 to December 2025, obtained by combining the light curve presented in \citet{GastaldelloSP} (shown in blue) with that in Fig.~\ref{fig:lc_2015_2026} (shown in black).}
    \label{fig:lc_2000_2026}
\end{figure*}

In Fig. \ref{fig:lc_2000_2026}, the resulting curve is combined with the long-term light curve previously generated by \citet{GastaldelloSP}.

\section{External data set}\label{sec3}

We complemented the analysis with two independent external tracers sensitive to charged-particle fluxes in near-Earth space: the high-energy reject rates measured by ACIS onboard \textit{Chandra} X-ray Observatory and the proton flux monitored by AMS-02. These datasets provide an external reference against which the temporal behaviour of the MOS2 outFOV count rate can be compared.

\subsection{\textit{Chandra}-ACIS reject rates}

\begin{figure*}
    \centering
    \includegraphics[width=0.8\textwidth]{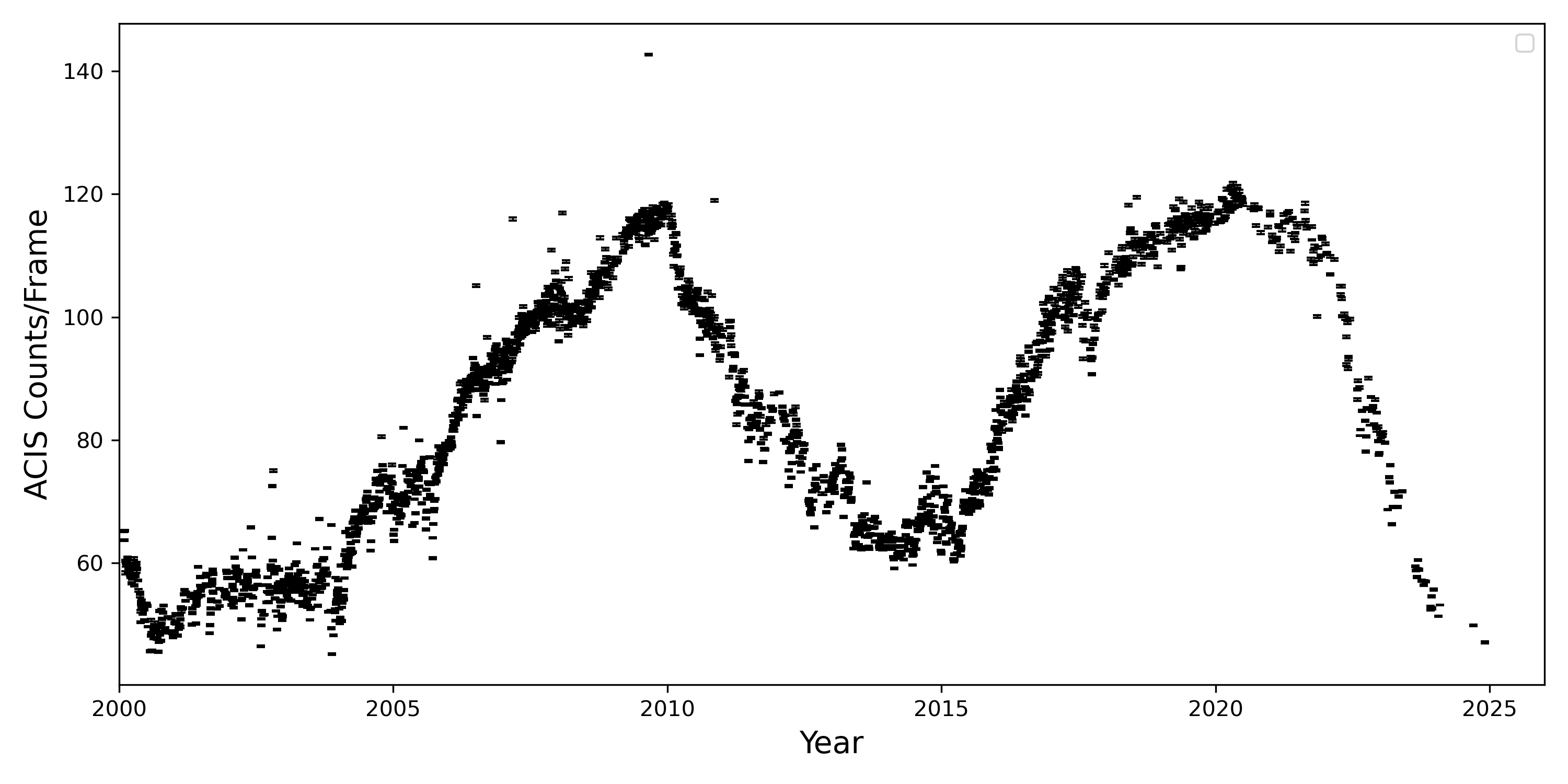}
    \caption{The high-energy rejected count rate observed by Chandra's ACIS-S3 CCD, plotted as a function of time from January 2000 through 2025, after filtering out SEP events. The data have been rebinned following the \textit{XMM-Newton} revolution intervals. Counts are expressed per full ACIS frame (3.241 s).}
    \label{fig:lc_chandra}
\end{figure*}

We included a particle-background proxy based on the Chandra X-ray Observatory ACIS reject rates \citep{Grant2024}. Like XMM EPIC, Chandra ACIS is a silicon-based detector in a HEO.  These data were taken while the instrument was stowed and not viewing the sky, analogous to the outFOV.  These reject rates correspond to events above the threshold (either 13 keV or 15 keV) recorded in the standard \textit{Chandra} telemetry products. Although such events are discarded on board, their rates are telemetered to the ground, providing a continuous monitor of the particle-induced background. The resulting light curve was filtered using the same SEP list described in Section~\ref{sec2} and subsequently binned according to the \textit{XMM-Newton} revolution intervals, the final result is shown in Fig. \ref{fig:lc_chandra}.

\subsection{AMS-02}

\begin{figure*}
    \centering
    \includegraphics[width=0.8\textwidth]{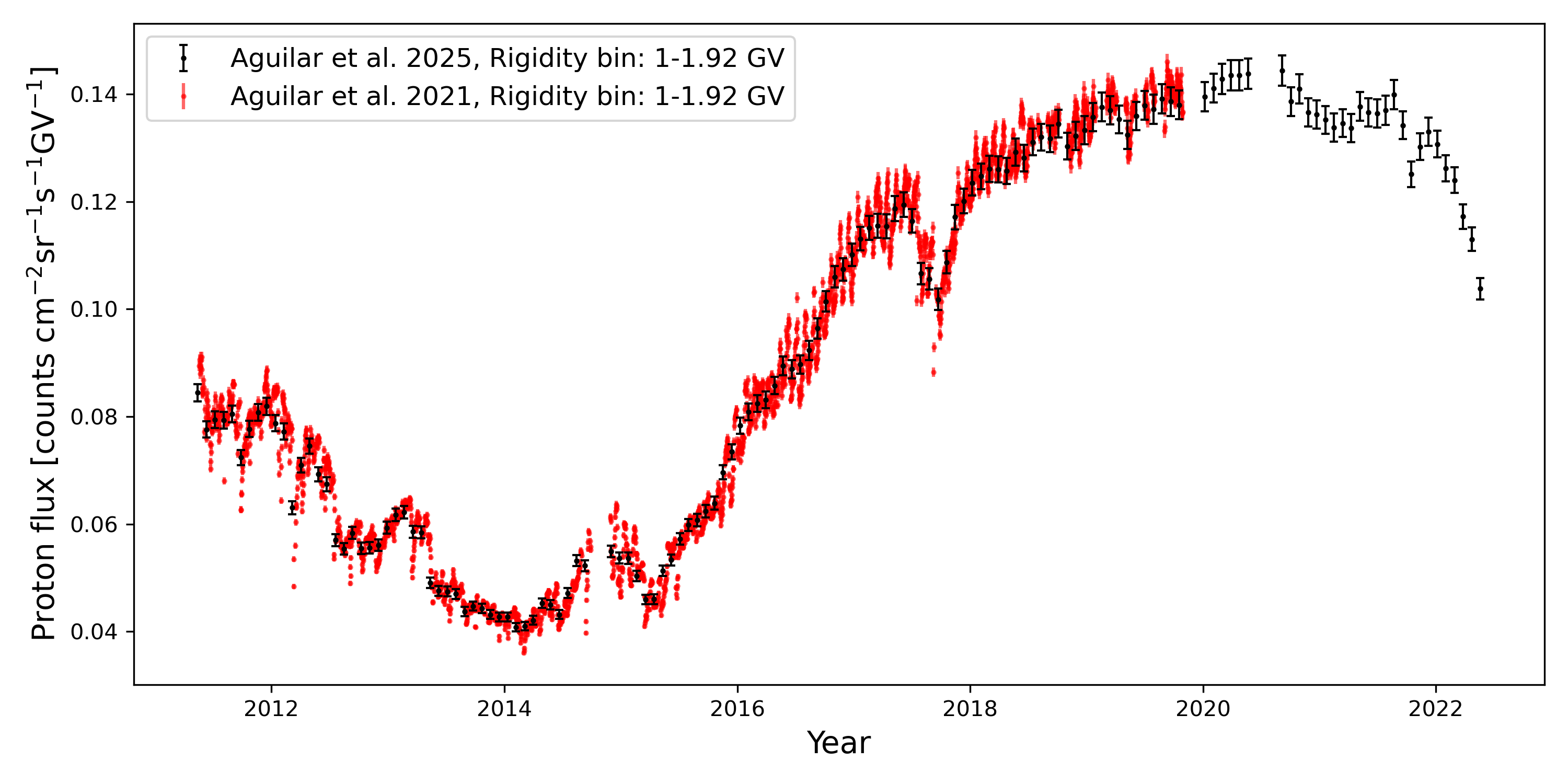}
    \caption{In red the daily proton flux published in \citet{Aguilar2021} from May 2011, to October 29, 2019. In black the proton flux binned over the Bartels rotation period  from May 2011 to June 2022, published in \citet{Aguilar2025}. Both datasets are in the rigidity bin between 1 and 1.92 GV.}
    \label{fig:lc_ams}
\end{figure*}

The Alpha Magnetic Spectrometer (AMS-02) is a particle physics experiment installed on the International Space Station in May 2011 aimed at exploring the high-energy cosmic rays and particles in the universe. In this work, we use proton flux measurements from the tables published by the AMS Collaboration in \citet{Aguilar2021} and \citet{Aguilar2025}. 
The first set of data (\citealt{Aguilar2021}, in red in Fig. \ref{fig:lc_ams}) contains daily-binned AMS proton flux data covering the period from May 20, 2011, to October 29, 2019. 
The second set \citep{Aguilar2025} extends up to June 2022, but is binned over the Bartels rotation period, i.e. 27-day intervals approximately corresponding to the synodic solar rotation period. 
For the data set from \citet{Aguilar2025} (in black in Fig. \ref{fig:lc_ams}), we selected the rigidity interval between 1 and 1.92 GV, corresponding to the low-rigidity regime where solar modulation effects are strongest and temporal variations in the proton flux are more pronounced. This low rigidity regime correspond to a proton energy range between 0.4 and 1.2 GeV.
For the data set from \citet{Aguilar2021}, smaller rigidity bins are available; therefore, to match the rigidity range adopted for the \citet{Aguilar2025} data set, we used the average flux of the first five rigidity bins.
Although AMS-02 operates in low-Earth orbit, the proton fluxes published by the AMS Collaboration are reconstructed to represent the cosmic-ray spectrum incident outside the Earth's magnetosphere, after accounting for the geomagnetic cutoff (with the fluxes error $<$ 2 \%, \citealt{Aguilar2021}) and detector response. Therefore, they can be directly compared with measurements probing the interplanetary particle environment.

\section{Results}\label{sec4}

We present here the results obtained from the analysis and comparison of the MOS2 outFOV count rate, the \textit{Chandra} X-ray Observatory ACIS reject rates, and the AMS-02 proton flux over their common time interval.

\subsection{Correlations}\label{subsec41}
As in \citet{GastaldelloSP}, we find that the temporal evolution of the MOS2 outFOV light curve closely follows that of both the \textit{Chandra} ACIS rejected rates and the AMS proton flux, suggesting a common dependence on the Galactic cosmic-ray environment. To quantify these relationships, we compared the MOS2 outFOV rate with both the \textit{Chandra} reject rate and the AMS proton flux, adopting a common temporal binning. We then modeled the correlations using a Bayesian linear regression framework \citep{Kelly2007}, incorporating measurement uncertainties in both variables as well as an intrinsic scatter term.
We adopted a relation of the form:
\begin{equation}
Y = \alpha + \beta X + \epsilon.
\end{equation}
The free parameters of the fit are the intercept $\alpha$, the slope $\beta$, and the intrinsic scatter $\epsilon$, which describes the dispersion around the linear relation. 
From this last parameter we derived a percentage scatter defined as the intrinsic scatter divided by the median of the y values ($0.0022$ counts/s/cm$^2$/keV). The estimate of the intrinsic scatter can be influenced by an overestimation of the uncertainties. Therefore, we also compute the root mean square deviation (RMSD), defined as
$
\mathrm{RMSD}=\sqrt{\frac{1}{N}\sum_{i=1}^{N}(Y_i-y_i)^2},
$
where $y_i$ is the observed value, $Y_i$ is the corresponding fitted value, and N is the number of data points. In a similar way we then derived a percentage value defined as the RMSD divided for the median of the y values, this value is used as an upper limit to the intrinsic value.
We implemented the regression using the Python package \textit{linmix} \citep{Kelly2007, Meyers2026}.

As in \citet{GastaldelloSP} we converted the MOS2 outFOV rate from counts s$^{-1}$, normalized to the FOV area, into counts s$^{-1}$ cm$^{-2}$ keV$^{-1}$, assuming an energy band width of 3.4 keV.

\begin{figure}
    \centering
    \includegraphics[width=0.6\textwidth]{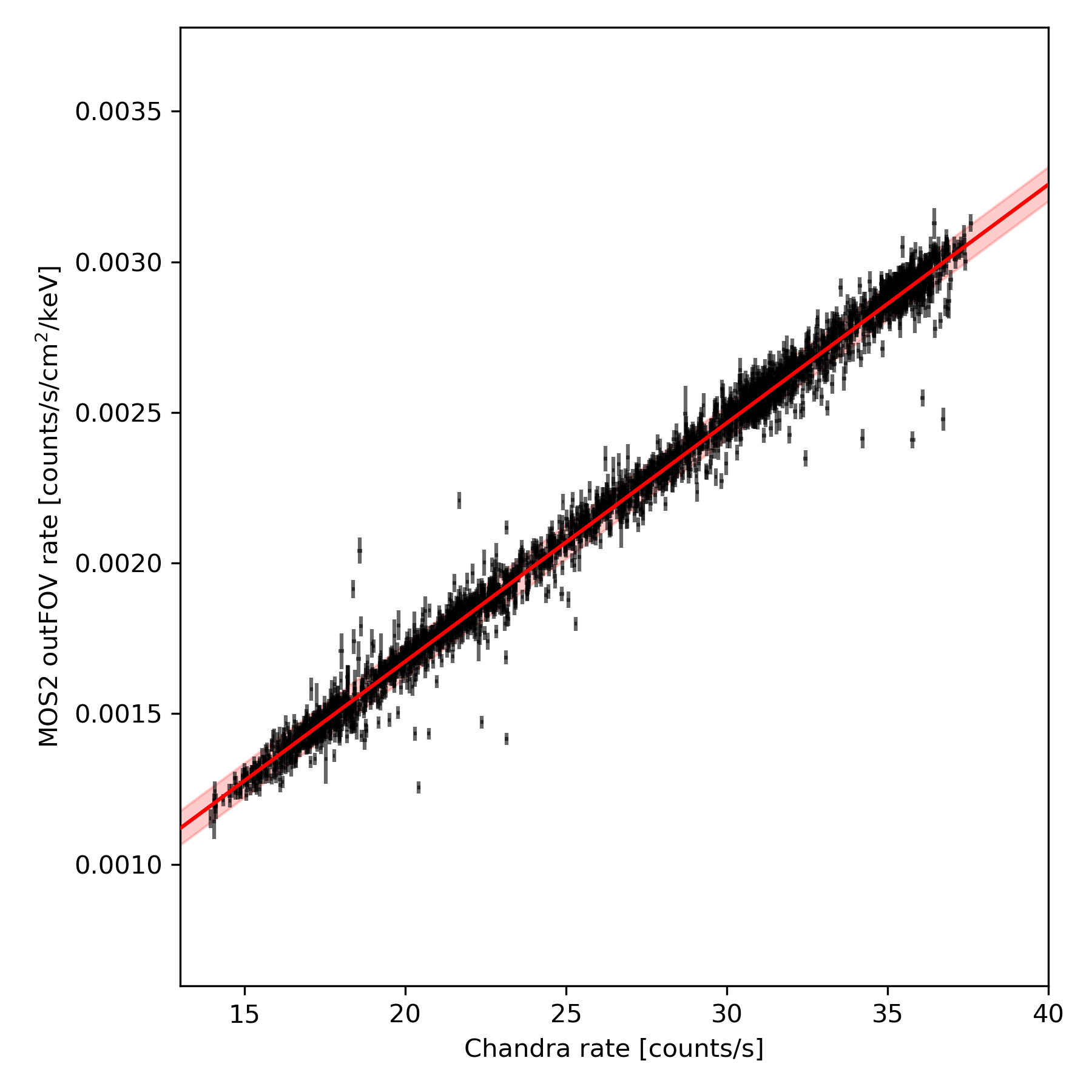}
    \caption{Correlation between the MOS2 outFOV count rate and the \textit{Chandra} ACIS-S3 reject rate over the 2000--2026 period, spanning two complete solar cycles. The red line shows the best-fit linear relation, while the shaded red band represents the upper limit derived from the RMSD.}
    \label{fig:XMMvsChandra}
\end{figure}

\begin{figure*}
    \centering
    \includegraphics[width=0.8\textwidth]{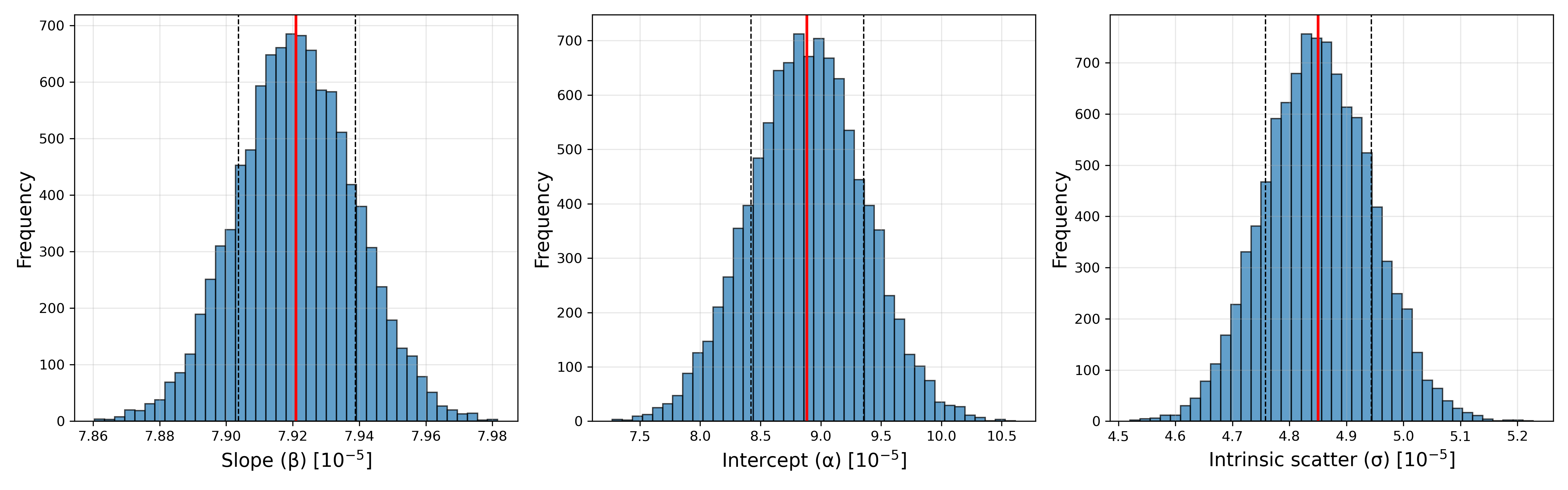}
    \caption{The samples of the posterior distributions of the three fitted parameters, derived from the correlation  between the MOS2 outFOV count rate and the \textit{Chandra} ACIS-S3 reject rate (Fig.~\ref{fig:XMMvsChandra}) using the \textit{linmix} Python code, are shown: the slope $\beta$ (left panel), the intercept $\alpha$ (central panel), and the intrinsic scatter $\sigma$ (right panel).}
    \label{fig:parameter_Chandra}
\end{figure*}

The correlation between the MOS2 outFOV rate and the \textit{Chandra} ACIS reject rate is shown in Fig.~\ref{fig:XMMvsChandra}, while the corresponding posterior distributions of the fit parameters are presented in Fig.~\ref{fig:parameter_Chandra}. Both datasets were previously binned according to the XMM-Newton revolution, as described in Sect. \ref{sec2} and \ref{sec3}. The inferred intrinsic scatter is very small, amounting to $2.25\pm0.04\%$\footnote{This value is expressed as a relative percentage and should not be confused with the intrinsic scatter shown in Fig. \ref{fig:parameter_Chandra}, which is reported in absolute units.}. The percentage RMSD provides an upper limit of 2.65\%.

\begin{figure}
    \centering
    \includegraphics[width=0.6\textwidth]{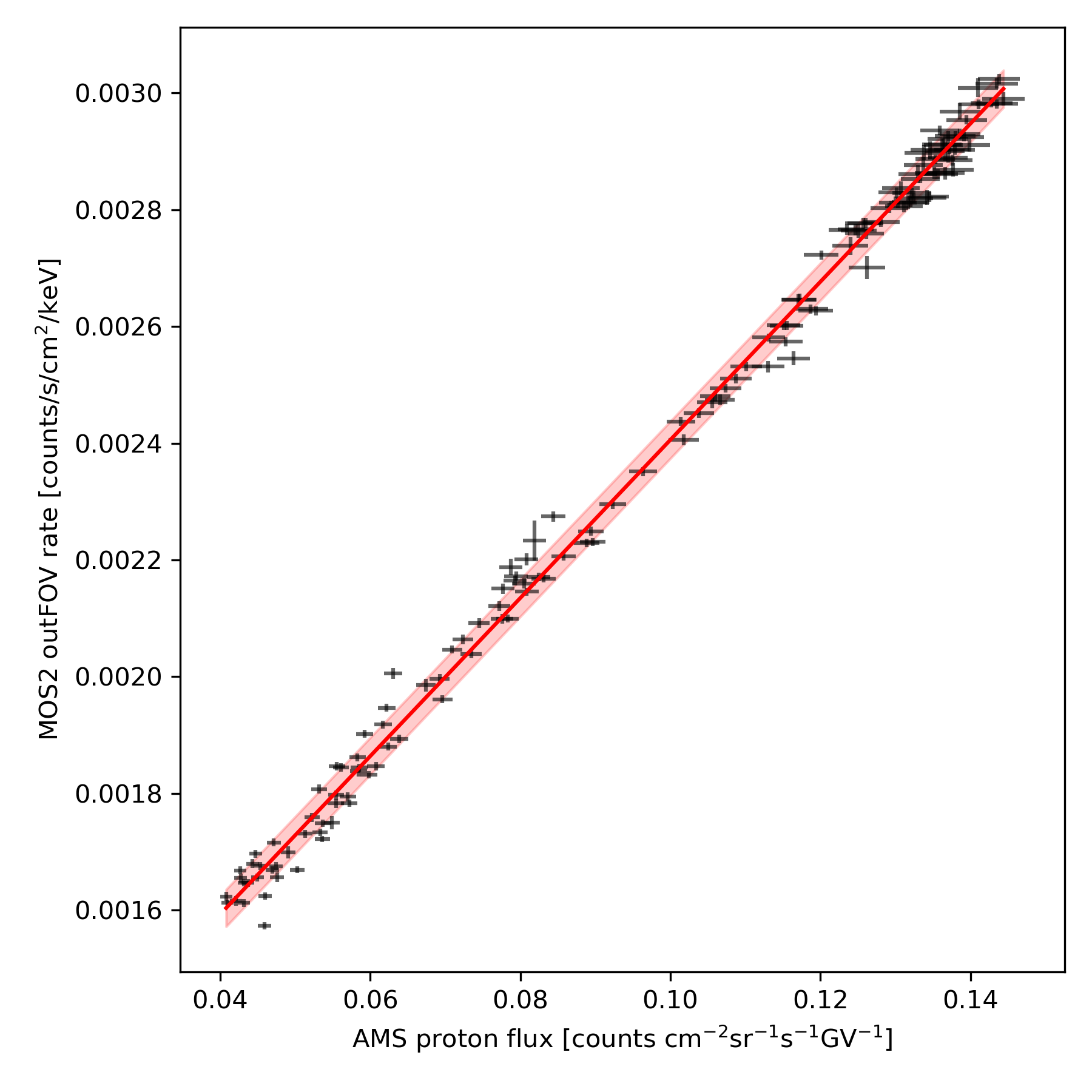}
    \caption{Correlation between the MOS2 outFOV count rate and the AMS proton flux. The relation is computed using data from 2011 to 2022, binned per Bartels rotation. The red line shows the best-fit linear relation, while the shaded red band represents the upper limit derived from the RMSD.}
    \label{fig:XMMvsAMS}
\end{figure}

\begin{figure*}
    \centering
    \includegraphics[width=0.8\textwidth]{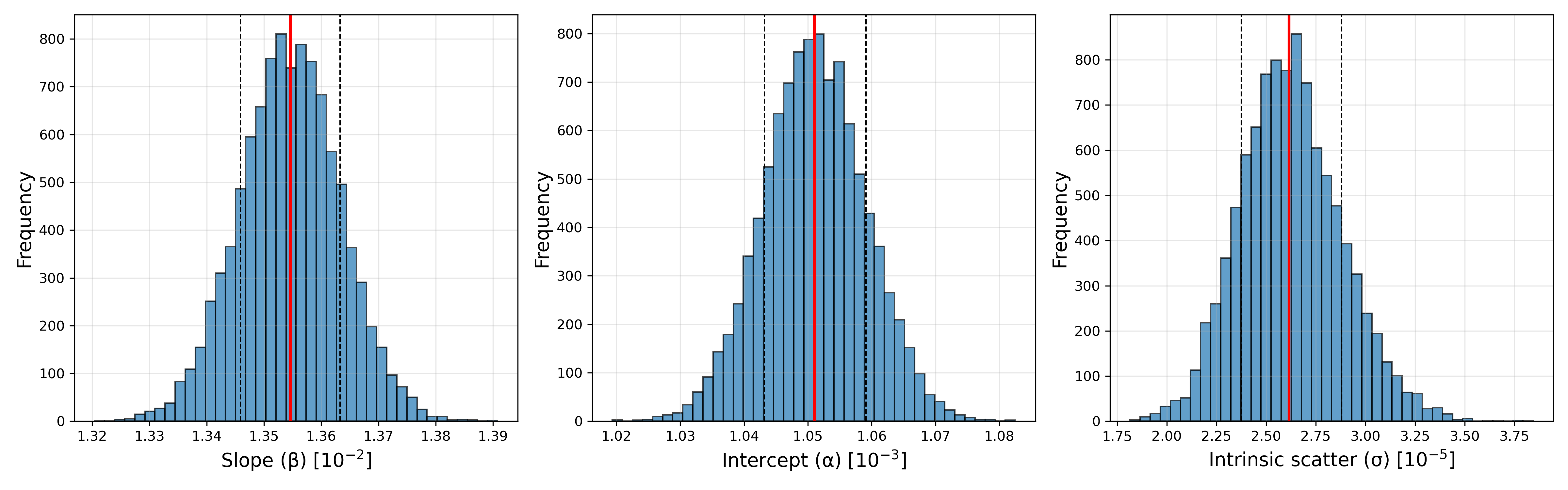}
    \caption{The samples of the posterior distributions of the three fitted parameters, derived from the correlation between the MOS2 outFOV count rate and the AMS proton flux (Fig.~\ref{fig:XMMvsAMS}) using the \textit{linmix} Python code, are shown: the slope $\beta$ (left panel), the intercept $\alpha$ (central panel), and the intrinsic scatter $\sigma$ (right panel).}
    \label{fig:parameter_AMS}
\end{figure*}

For the comparison with the AMS proton flux, we note that AMS data are available up to 2022 only in bins corresponding to Bartels rotations. Therefore, before performing the correlation analysis, we rebinned the MOS2 outFOV data over the same Bartels-rotation intervals, in order to exploit the maximum temporal range currently available from the published AMS dataset. The correlation between the MOS2 outFOV rate and the AMS proton flux is shown in Fig.~\ref{fig:XMMvsAMS}, while the corresponding posterior distributions of the fit parameters are presented in Fig.~\ref{fig:parameter_AMS}. The inferred intrinsic scatter is even smaller than the previous relation, amounting to $1.1 \pm 0.1$\%. With an upper limit due to the percentage RMSD of 1.34.

\subsection{Time lag}

\begin{figure*}
    \centering
    \includegraphics[width=0.8\textwidth]{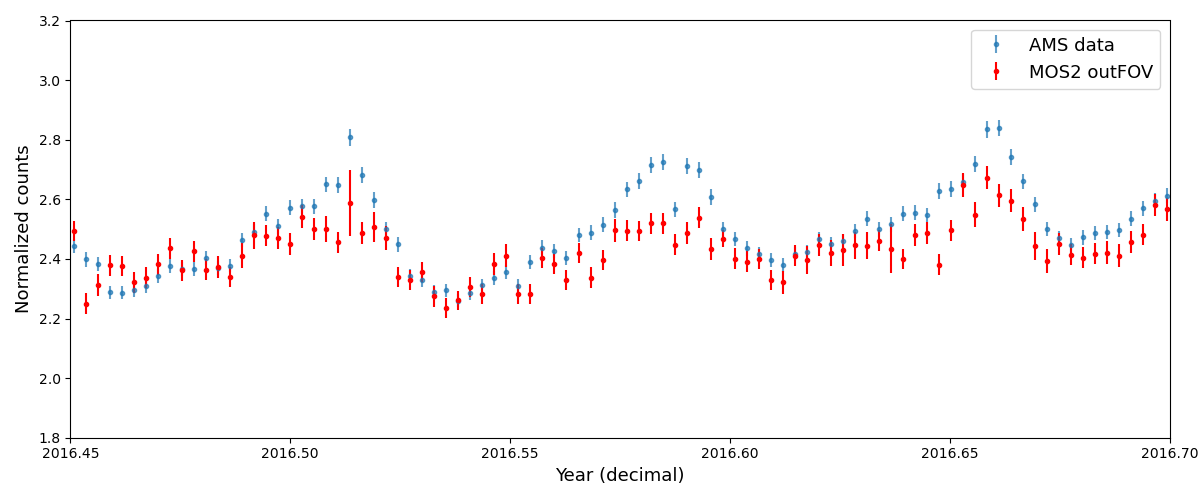}
    \caption{Comparison between the daily-binned AMS proton flux measurements \citep{Aguilar2021} and the daily-binned MOS2 outFOV count rate from June to September 2016. Both datasets are rescaled to arbitrary units for visual comparison.}
    \label{fig:AMS_XMM_2016}
\end{figure*}

\begin{figure}
    \centering
     \includegraphics[width=0.6\textwidth]{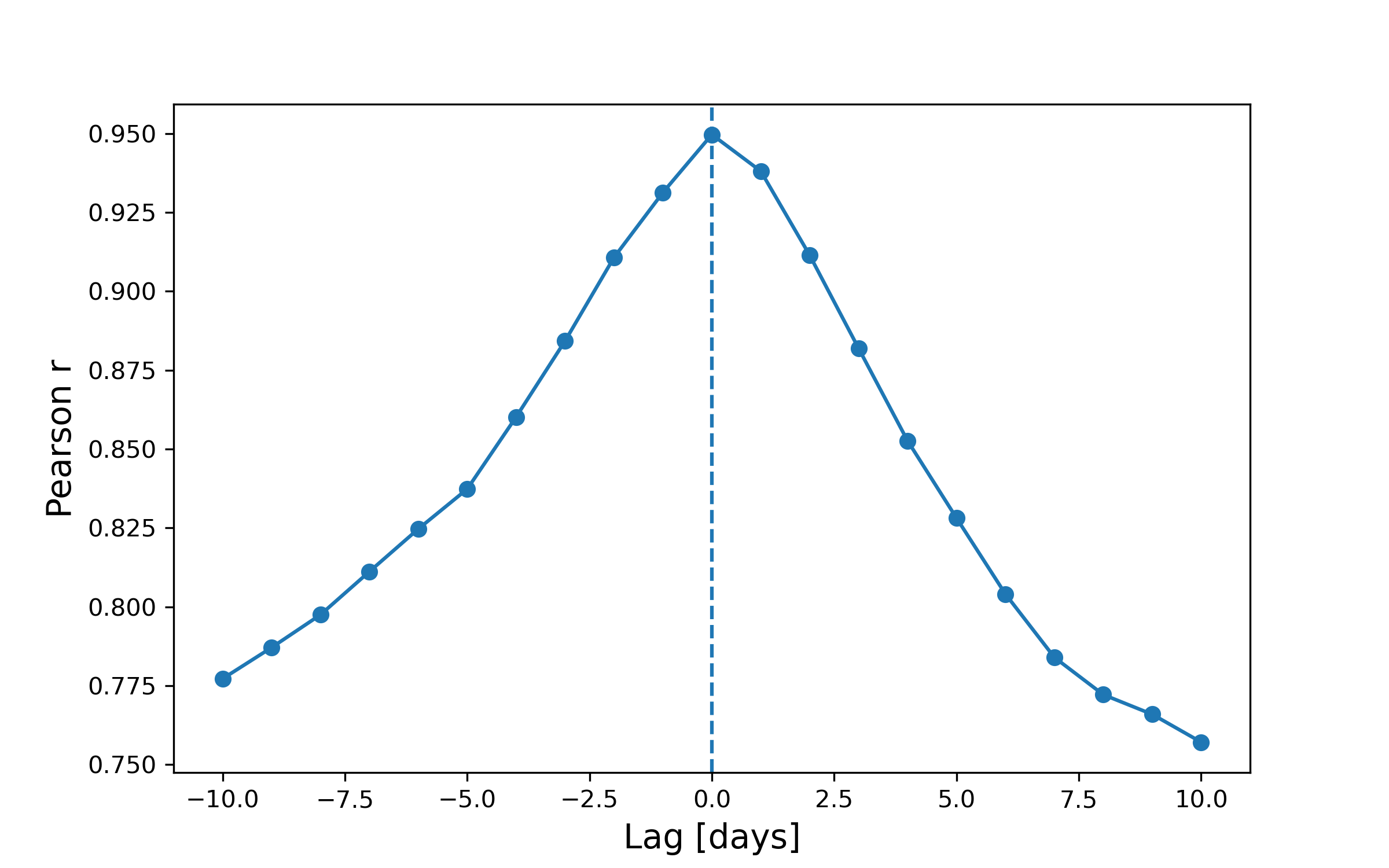}
    \caption{Correlation coefficient as a function of time lag between the normalized AMS proton flux and the MOS2 outFOV count rate in the 2016--2017 interval, during which \citet{Sarkar2024} reported evidence for a time lag. Positive lags indicate that the \textit{XMM-Newton} outFOV signal is delayed with respect to the AMS proton flux, while negative lags indicate the opposite. The maximum absolute Pearson correlation coefficient identifies the lag at which the two time series are best aligned.}
    \label{fig:lag_XMM_AMS}
\end{figure}

In \citet{Sarkar2024}, a time lag of $\simeq$ 6 days between the AMS proton flux and the ACIS/EPIC-pn reject rates was reported for the 2016--2017 period. We therefore examined whether a comparable delay is also present in the MOS2 outFOV rate in the same period. For this analysis, we used the AMS proton flux data binned on a daily timescale \citep{Aguilar2021} and rebinned the MOS2 outFOV light curve accordingly, in order to probe possible delays as short as one day. A close-up of the data presented in Fig.~\ref{fig:AMS_XMM_2016} already indicates the absence of such a lag in our dataset. To further test this, we systematically shifted the MOS2 daily-binned light curve relative to the AMS daily-binned proton flux and computed the resulting correlation for a range of temporal offsets. The results, shown in Fig.~\ref{fig:lag_XMM_AMS}, demonstrate that the correlation is clearly maximized at zero time lag, with a Pearson correlation coefficient of $r=0.95$.
We repeated the same analysis using the \textit{Chandra} ACIS reject-rate data described in Sect. \ref{sec3}, also rebinned on a daily timescale, and again found no statistically significant time lag with respect to the AMS proton flux.
The same conclusion was independently reached by \citet{Schellenberger2025}, who compared the primaries in PN SWM data with the \textit{Chandra} HRC anti-coincidence shield rates and found no significant time lag between the two datasets.

\subsection{Periodicity in the data}

\begin{figure}
    \centering
    \includegraphics[width=0.6\textwidth]{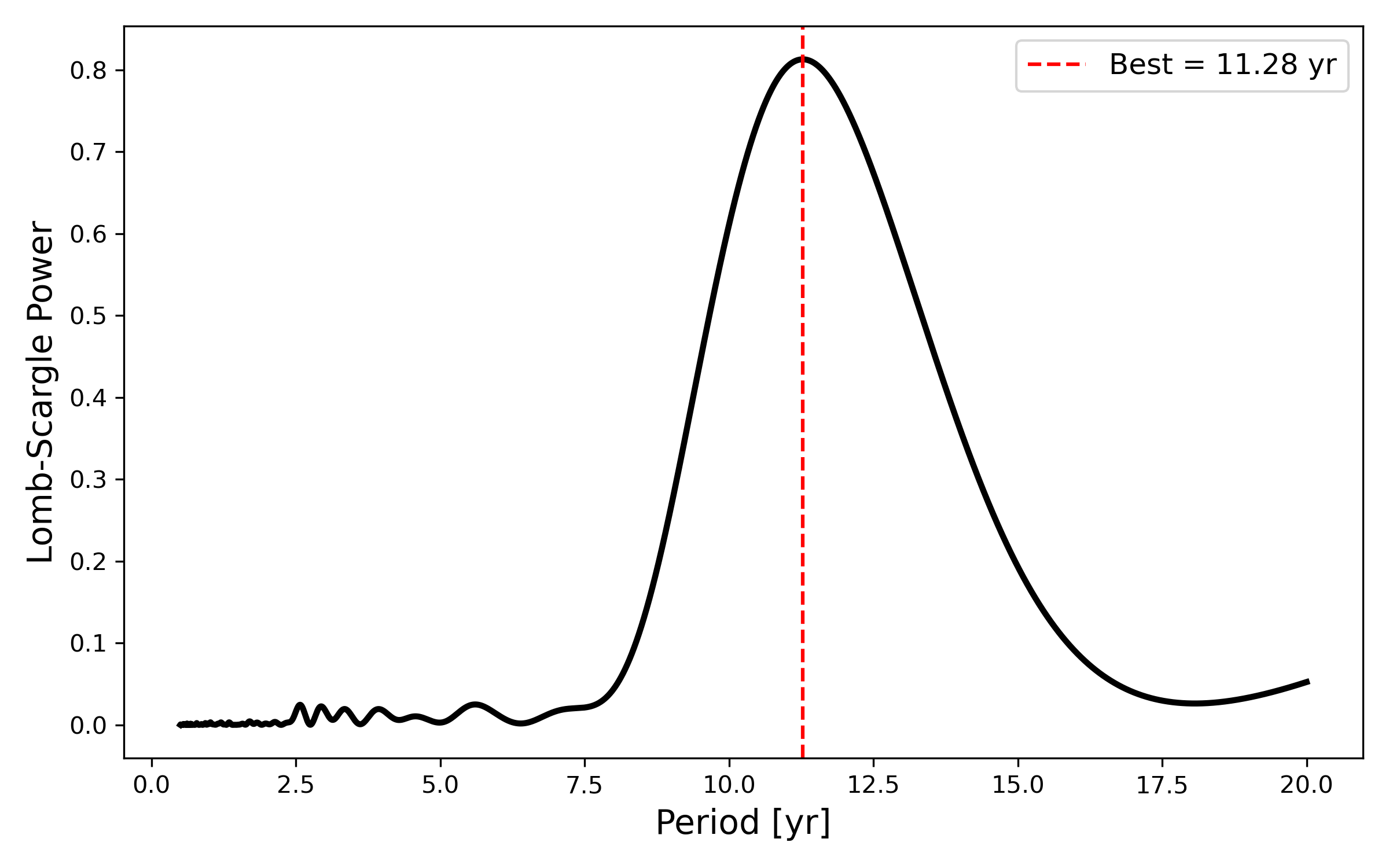}
    \caption{Lomb--Scargle periodogram of the MOS2 outFOV light curve. A strong and dominant peak is detected at a period of $11.28$ yr (red dashed line).}
    \label{fig:Lomb}
\end{figure}

A modulation of the MOS2 outFOV light curve at the $\sim$11-year solar cycle timescale is clearly visible in the data shown in Fig. \ref{fig:lc_2000_2026}. This periodicity is confirmed by the Lomb--Scargle power spectrum\footnote{The Lomb--Scargle periodogram extends Fourier spectral analysis to unevenly sampled time series.} \citep{VanderPlas2018} in Fig. \ref{fig:Lomb}, where it appears as the dominant and highly significant peak in the periodogram.

\begin{figure}
    \centering
    \includegraphics[width=0.6\textwidth]{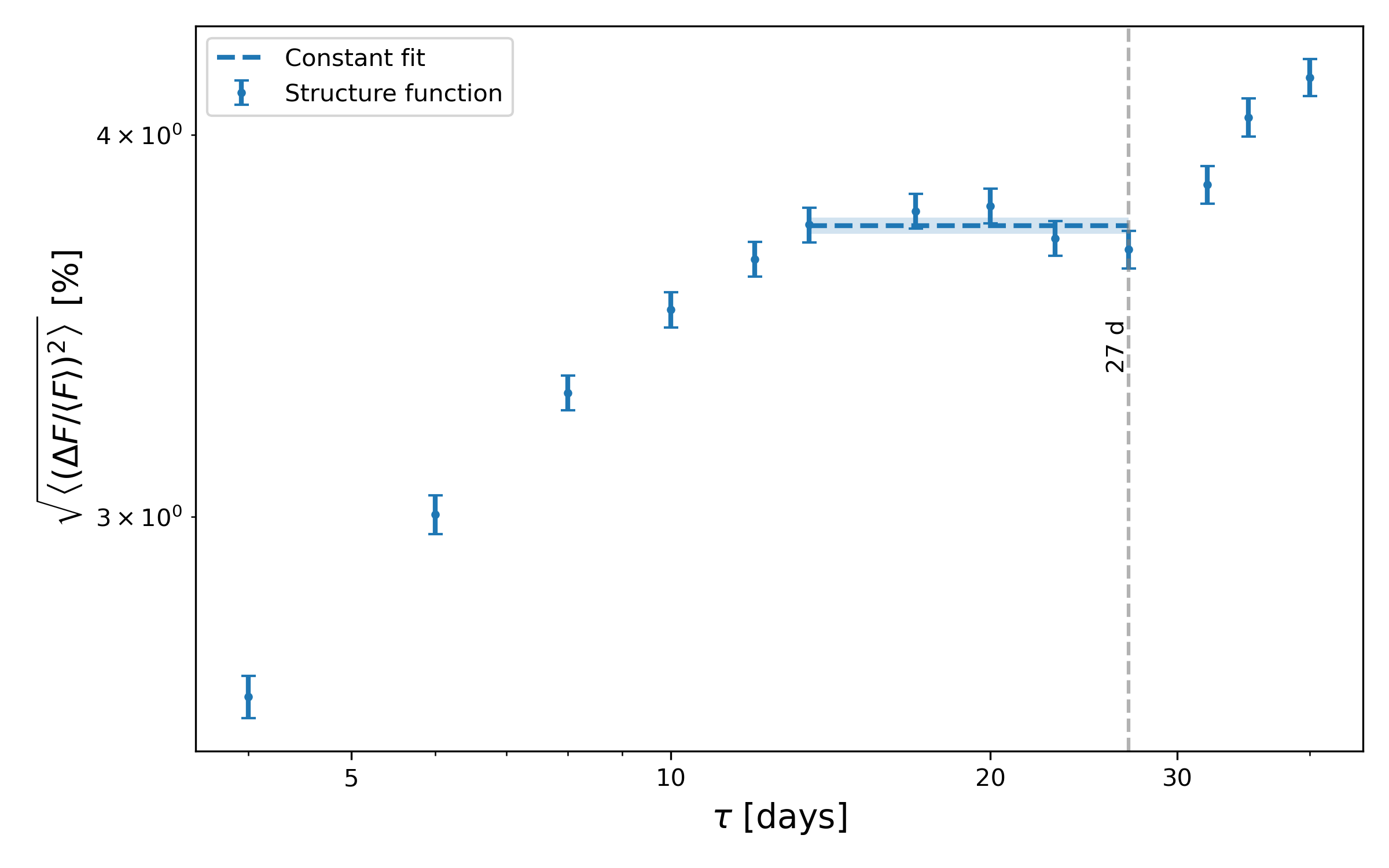}
    \caption{Structure function of the \textit{XMM-Newton} outFOV count-rate time series, computed over timescales between 4 and 40 days. The plotted quantity represents the fractional root-mean-square variability, $\sqrt{\langle (\Delta F / \langle F \rangle)^2 \rangle}$, expressed in percent as a function of the time lag $\tau$.}
    \label{structure}
\end{figure}

In this work, as in several recent studies (e.g., \citealt{Aguilar2021}; \citealt{Sarkar2024}; \citealt{Schellenberger2025}), we also investigate variability on shorter timescales.
To investigate such small-scale periodicities, we first computed the structure function \citep{Dewit2013} of the \textit{XMM-Newton} outFOV count-rate data over the 2000--2025 interval, focusing on timescales between 4 and 40 days (Fig. \ref{structure}).
The adopted time lags were selected taking into account the 2-day binning of the MOS2 light curve and to adequately sample the range of timescales around the expected $\sim$27-day Solar rotation period and its harmonics. These timescales were therefore defined a priori on the basis of the temporal resolution of the data and the physical timescales of interest.
The structure function quantifies the fractional variability of the outFOV count rate, F, over different timescales and is particularly well suited for identifying characteristic variability timescales in unevenly sampled datasets. In the presence of a periodic signal, the structure function is expected to exhibit a plateau at separations corresponding to the characteristic period, reflecting the recurrence of similar count-rate values.        
In our case, a flattening of the structure function is observed on timescales approaching the expected $\sim$27-day Solar rotation period. A constant fit to the plateau is statistically acceptable ($\chi^2/\rm{d.o.f.}=1.03$, p-value=0.39). Given the adopted sampling in time lag, the structure function is not intended to provide a precise measurement of the endpoint of the plateau at 27 days, but rather to show that the characteristic variability timescale is consistent with the Solar rotation period.

\begin{figure*}
    \centering
    \includegraphics[width=1.\textwidth]{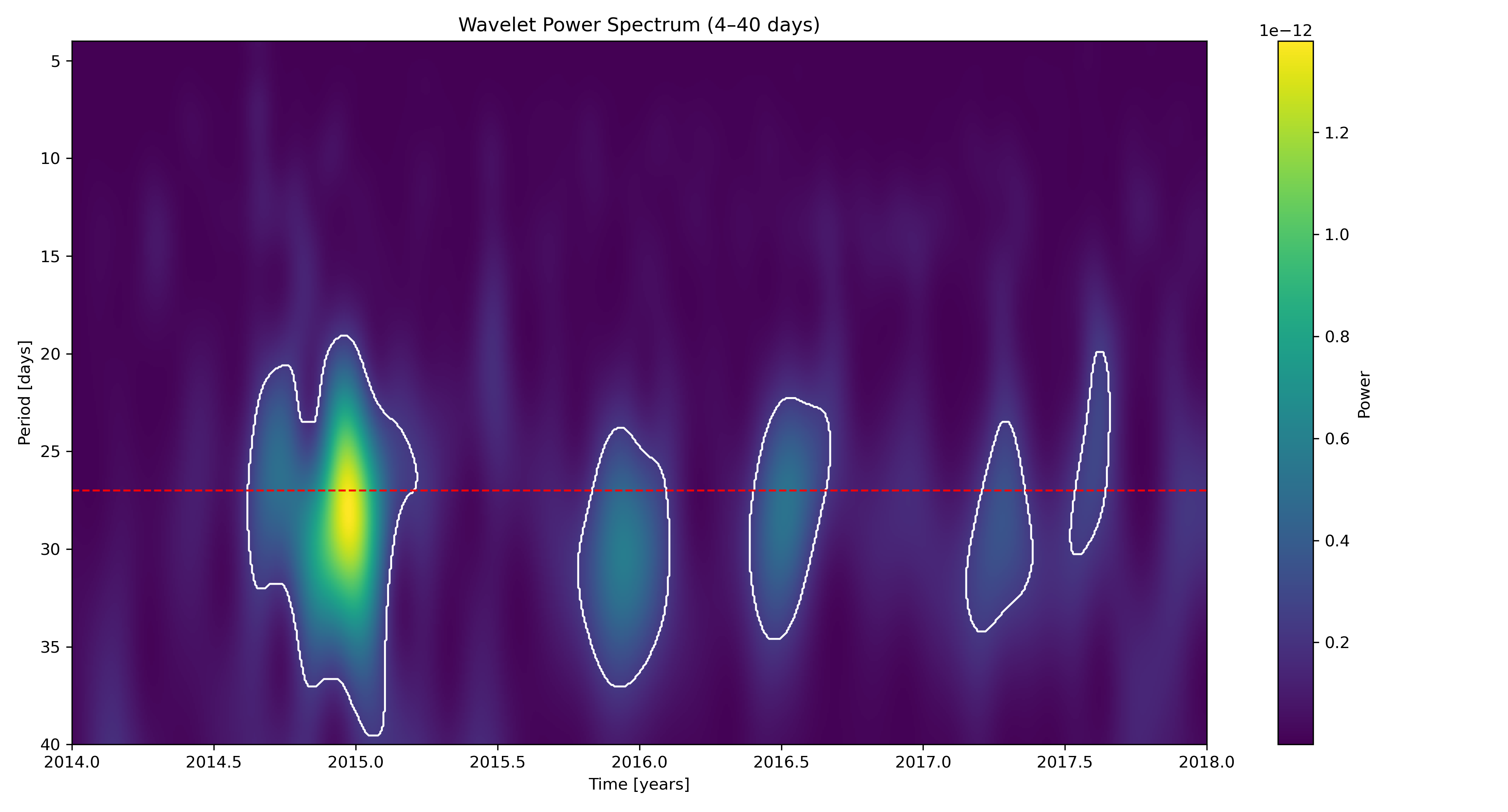}
    \caption{Wavelet power spectrum of the \textit{XMM-Newton} EPIC outFOV count-rate data for the 2014--2018 interval. The enhanced power around 2015 coincides with the maximum phase of Solar Cycle 24. The horizontal red dashed lines indicate the 27 day periodicity. White contours mark the 95\% confidence level.}
    \label{wavel_spectra2014_2018}
\end{figure*}

We also performed a wavelet time--frequency analysis on the \textit{XMM-Newton} data, focusing on the same range of periodicities in order to investigate whether these signals persist over the full temporal baseline or are instead confined to specific epochs. Unlike the structure function, which provides an average characterization of variability as a function of time lag, the wavelet transform allows the temporal evolution of periodic signals to be tracked simultaneously in both time and frequency domains. The wavelet transform was computed using a Morlet mother wavelet, as implemented in the PyWavelets package \citep{Lee2019}, and regions exceeding the 95\% confidence level are highlighted in white in the wavelet power spectra. A clear and statistically significant quasi-periodicity at $\sim$27 days is detected also in this analysis, see the red line in Fig.~\ref{wavel_spectra2014_2018} over plotted as reference. This signal is present throughout most of the considered time interval ($\sim80\%$), except during solar minimum conditions, when the modulation appears significantly weaker, in good agreement with the results reported in \citet{Gil2017}. In contrast, the signal is particularly pronounced during the 2014--2018 period (Fig.~\ref{wavel_spectra2014_2018}), in agreement with previous findings reported by \citet{Aguilar2021} and \citet{Sarkar2024}. We also explicitly searched for signals at $\sim$13.5 and $\sim$9 days, corresponding to the second and third harmonics of the $\sim$27-day Solar rotation period, respectively. Such harmonics may arise from recurrent heliospheric structures at different Solar longitudes, which can produce multiple modulations within a single Solar rotation (e.g. \citealt{Gil2017}). However, no statistically significant power is detected at either of these shorter timescales.

\section{Discussion}\label{sec5}

In this section, we discuss the implications of the temporal variability observed in the MOS2 outFOV background over 25 years of data and its connection with the modulation of Galactic Cosmic Rays over the solar cycle, particularly in the context of future missions such as NewAthena.

\subsection{Correlation with the Galactic Cosmic Rays}\label{sec5.1}

The correlation analysis presented in Sect.~\ref{subsec41} provides strong evidence that the MOS2 outFOV count rate is closely connected to the Galactic cosmic-ray environment, confirming the results previously reported by \citet{GastaldelloSP}. In particular, the very small intrinsic scatter derived from the Bayesian regression with the AMS proton flux (Fig. \ref{fig:XMMvsAMS}, $1.1 \pm 0.1$ \%), which directly probes the proton component of Galactic Cosmic Rays in low-Earth orbit, strongly supports the interpretation that the unfocused particle background measured by \textit{XMM-Newton} is primarily driven by variations in the incident cosmic-ray flux. The best-fit intercept is $\sim1.05\times10^{-3}$ counts s$^{-1}$ cm$^{-2}$ keV$^{-1}$, corresponding to about one-third of the mean MOS2 outFOV count rate ($\sim2\times10^{-3}$ counts s$^{-1}$ cm$^{-2}$ keV$^{-1}$). This indicates the presence of a substantial background component that remains approximately constant over the solar cycle. During solar maximum, when the Galactic cosmic-ray component is most strongly suppressed, this constant component accounts for up to $\sim60\%$ of the observed MOS2 outFOV count rate.
Furthermore, a comparable strong correlation between the MOS2 outFOV count rate and the \textit{Chandra}-ACIS reject rates (Fig. \ref{fig:XMMvsChandra}, $2.25 \pm 0.04$\%) can naturally be explained by the common dependence of both detectors on the flux of high-energy Galactic protons. The outliers in the correlation can be attributed to residual SEP contamination, as reported by \citet{Gastaldello2017,GastaldelloSP}. The slope derived from our correlation analysis is fully consistent with the value previously reported by \citet{GastaldelloSP}, supporting not only the common dependence described above, but even a stability of this relation over an extended temporal range.

In \citet{GastaldelloSP}, the correlation between the MOS2 outFOV background and the EPIC Radiation Monitor (ERM) rates was found to be significantly weaker than the correlation found between the MOS2 outFOV count rate and data from the Electron Proton Helium INstrument (EPHIN) onboard the Solar and Heliospheric Observatory (SOHO). This difference was interpreted as a consequence of the different particle energies ranges sampled by the instruments. The ERM onboard \textit{XMM-Newton} includes a High Energy (HE) unit, and in \citet{GastaldelloSP} the authors used the HES0 rate, which is sensitive to protons in the 8 -- 40 MeV energy range, significantly lower than the 250 MeV -- 1.6 GeV range covered by SOHO/EPHIN. The very tight correlation we find with the AMS proton flux, which also probes high-energy Galactic Cosmic Rays (400 MeV -- 1.2 GeV), further supports this interpretation. 

\begin{figure}
    \centering
    \includegraphics[width=0.6\textwidth]{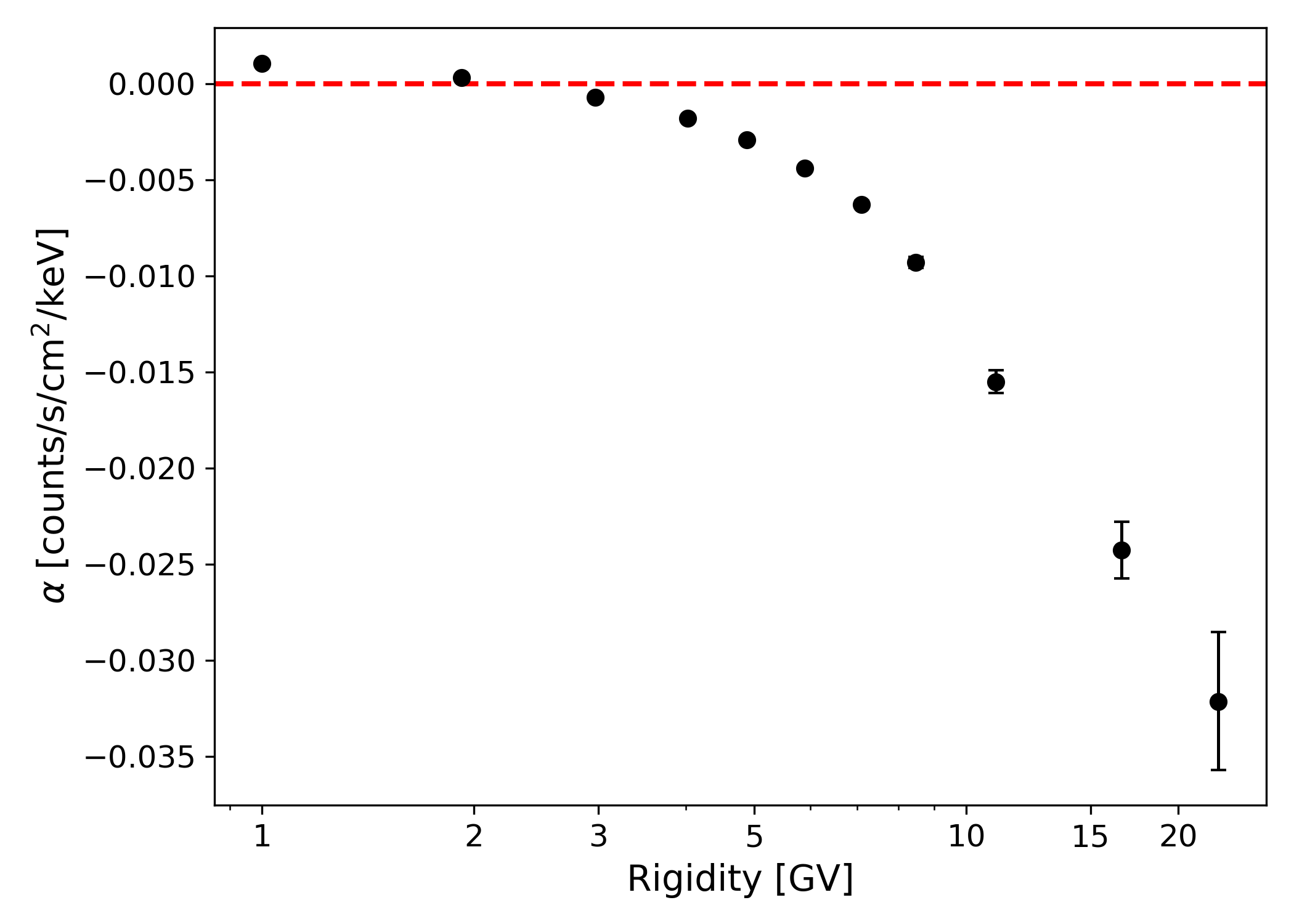}
    \caption{Dependence on proton rigidity of the best-fit intercept $\alpha$ of the relation between the MOS2 outFOV count rate and the AMS--02 proton flux. The red dashed line marks $\alpha=0$.}
    \label{alphavsrigbin}
\end{figure}

To identify the proton population most closely associated with the unfocused background, we repeated the correlation analysis using AMS proton fluxes measured in different rigidity intervals. We find that the strongest positive correlation is obtained in the lowest rigidity intervals (corresponding to 400 MeV -- 2.18 GeV protons). Starting from the third rigidity bin, the constant term ($\alpha$) becomes negative (see Fig. \ref{alphavsrigbin}).

This behavior is expected because, at higher rigidities, the amplitude of solar modulation decreases substantially. In the rigidity range above a few GV, the proton flux varies by less than $\sim$ 10\% over the solar cycle, making it a poor tracer of the long-term modulation observed in the MOS2 unfocused background. As a consequence, the correlation becomes progressively weaker and the fitted slope is no longer physically meaningful.

Consequently, instruments sensitive to higher-energy protons than ERM, such as EPHIN or AMS, provide a more reliable tracer of the unfocused background. This result is particularly relevant for future missions such as NewAthena, where monitoring the high-energy particle environment will be crucial for accurately characterizing and mitigating the instrumental background. 

\subsection{The Hale cycle}

The long-term modulation of cosmic rays, strongly influenced by the $\sim$22-year solar magnetic polarity (Hale) cycle, has been widely recognized and studied for several decades and is now considered a well-established feature of cosmic-ray variability (see \citealt{McDonald2010} and references therein). The transport of Galactic Cosmic Rays from the interstellar medium into the inner heliosphere is strongly affected by the large-scale solar wind and by the turbulent interplanetary magnetic field, a combined process commonly referred to as \textit{solar modulation}. In particular, periods of high solar activity correspond to a stronger modulation of cosmic rays, resulting in a reduced GCR flux reaching the inner heliosphere, whereas during solar minima the weaker heliospheric shielding allows a larger flux of particles to penetrate into the Solar System. Since the unfocused particle background detected by X-ray observatories is largely produced by cosmic rays interacting with the spacecraft and detector structures (as described in \ref{sec5.1}), a corresponding long-term modulation is expected in the unfocused background rates. This modulation is characterized by a double-peaked profile, with one sharp peak and another exhibiting a flat-topped shape, as clearly seen in Fig.~\ref{fig:lc_2000_2026} and Fig.~\ref{fig:lc_chandra}.
The distinct morphology of these modulation cycles is due to the solar magnetic field reverse polarity every $\sim$11 years, causing the cosmic rays to reach the earth from different heliosphere directions. During A$>$0  polarity cycles, where (A) denotes the polarity of the large-scale heliospheric magnetic field, positively charged cosmic rays drift into the inner heliosphere mainly through the polar regions, resulting in the observed flat-topped intensity profiles \citep{Potgieter2013}. 
Conversely, during A$<$0 cycles, cosmic rays mainly enter the inner heliosphere by following the heliospheric current sheet. In this configuration, the observed cosmic-ray flux becomes much more sensitive to changes in the shape and inclination of the heliospheric current sheet \citep{Potgieter2013}. As a result, the temporal profiles of the flux tend to show the characteristic sharp-peaked structure observed during A$<$0 epochs.

\subsection{27 days periodicity}

A likely explanation for the variability associated with the $\sim$27-day solar rotation period is the passage of corotating interaction regions, which originate from the interaction between fast and slow solar-wind streams, often associated with long-lived coronal holes. As the Sun rotates, these structures recurrently propagate through the heliosphere, producing periodic variations in the solar wind and interplanetary magnetic field that modulate the propagation of Galactic Cosmic Rays and, consequently, the particle flux reaching the inner Solar System. This provides a natural explanation for the $\sim$27-day variability observed in the MOS2 outFOV background. Large-scale drift effects, as those responsible for the dependence of cosmic-ray modulation on the $\sim$22-year Hale cycle, may also contribute to the observed variability \citep[e.g.][]{Modzelewska2021}.
The absence of significant periodicities at $\sim$13.5 and $\sim$9 days may indicate that the particles responsible for the MOS2 unfocused background are not those for which these higher-order harmonics are most prominent. In \citet{Aguilar2021}, the authors showed that the strength of the different periodicities depends on proton rigidity, with the $\sim$9-day and $\sim$13.5-day modulations becoming more pronounced at rigidities of $\sim$10$-$20 GV and then decreasing toward $\sim$100 GV. This interpretation is further supported by the fact that our data do not show a strong correlation with proton populations in these rigidity ranges.

 The presence of a recurrent $\sim$27-day modulation also provides a characteristic timescale for future background-monitoring strategies. In order to properly sample this variability, particle monitors onboard future missions such as \textit{NewAthena} should provide measurements on timescales significantly shorter than the solar rotation period, to know the \textit{NewAthena} background with a precision of 2\%. 
Assuming a modulation amplitude of $\sim$16\% over a 27-day period (which corresponds to the maximum variability observed over the 25-year period analyzed), this would require sampling the background at least $\sim$8 times per cycle, corresponding to a cadence of about 3--4 days. Such a cadence would allow the long-term particle background variations associated with solar rotation to be properly tracked and corrected.
Such monitoring could be achieved either through dedicated closed observations or by means of an onboard particle monitor continuously tracking the high-energy particle environment.

\subsection{Compton Scattering of Cosmic X-Ray Background Photons}

The correlation between the MOS2 outFOV count rate and the AMS proton flux shows evidence for a residual constant component, represented by the intercept term of the linear relation shown in Fig.~\ref{fig:XMMvsAMS}. We interpret this as evidence for a contribution to the unfocused background that is distinct from the Galactic cosmic-ray component directly traced by AMS. Similar results were previously reported by \citet{Marelli2021}, using the \textit{XMM-Newton} pn detector and SOHO/EPHIN data, and by \citet{GastaldelloSP}, using the \textit{XMM-Newton} MOS outFOV and SOHO/EPHIN data.

As discussed by \citet{Marelli2021} and \citet{GastaldelloSP}, this component is unlikely to be an artifact of the analysis. In addition, its approximately constant intensity over the solar cycle argues against an origin associated with Galactic cosmic rays or long-term activation processes. A possible interpretation is that this residual component originates from high-energy X-ray photons undergoing Compton scattering interactions within the detector and surrounding structures.
As in the previously cited papers we estimated an upper and lower bound to this component repeating the correlation between MOS2 and AMS data, taking advantage of the finer rigidity bins available in the daily AMS data \citep{Aguilar2021}. The upper bound was derived using the AMS data in the rigidity bin between 1 and  1.16 GV (0.4-0.5 GeV) and correspond to $1.2\times10^{-3}\,\rm{cts\,s^{-1}\,cm^{-2}\,keV^{-1}}$. While the lower bound taking the AMS data in the rigidity bin between 1.71 and  1.92 GV (1.0-1.2 GeV) and correspond to $6.8\times10^{-4}\,\rm{cts\,s^{-1}\,cm^{-2}\,keV^{-1}}$.

\section{Summary}\label{sec6}

The main results of this work can be summarized as follows:

\begin{itemize}
    \item The MOS2 outFOV count rate exhibits a strong long-term modulation that closely follows the evolution of Galactic cosmic rays over two complete solar cycles. The observed temporal changes are fully consistent with the expected solar modulation of GCRs.
    \item We find an exceptionally tight correlation between the MOS2 outFOV background and the AMS-02 proton flux. The inferred intrinsic scatter is only $1.1\pm0.1\%$, a strong correlation is also observed with the Chandra ACIS reject rates ($2.25\pm0.04\%$).
    \item A Lomb--Scargle periodogram confirms the presence of a highly significant periodicity at $\sim$ 11 yr related to the solar cycle. At shorter timescales, both the structure-function and wavelet analyses reveal a persistent $\sim27$-day modulation associated with solar rotation. No statistically significant periodicities are detected at $\sim13.5$ or $\sim9$ days.
    \item The correlation with AMS data confirms the presence of a residual background component that does not scale with the Galactic cosmic-ray flux. Consistent with previous studies, this component is likely unrelated to cosmic-ray modulation and may originate from high-energy X-ray photons undergoing Compton scattering within the detector and surrounding structures.
    
\end{itemize}

This finding has direct implications for the design of background-monitoring and mitigation strategies for future missions such as \textit{NewAthena}. In particular, the observed $\sim27$-day modulation suggests that continuous monitoring of the high-energy particle environment, with a cadence of 3-4 days, will be sufficient to track the dominant short-term variations of the instrumental particle background and achieve background predictions at the $\sim2\%$ level set as one of the mission requirements.

More generally, the study of GCR-induced backgrounds in X-ray detectors can be placed within the broader effort to characterize and understand the global modulation of Galactic cosmic rays throughout the heliosphere \citep[e.g.][]{Rankin2022}. The complex spatial and temporal variability of the heliospheric particle environment makes direct in-situ measurements particularly important for achieving the accuracy and precision required for background characterization. In this context, measurements provided by AHEPaM will allow the incident particle environment experienced by \textit{NewAthena} to be monitored directly.
Such measurements are also essential for establishing the relation between the incident GCR flux and the particle background actually measured by the X-ray detectors. This relation depends on the interaction of cosmic rays with the spacecraft and detector structures and therefore should not rely uniquely  from the measurement of another, not in situ, particle monitor. Increasingly sophisticated Geant4 simulations provide a powerful framework for modelling these interactions and the resulting instrumental background. Direct in-situ particle measurements will provide the complementary information needed to constrain and validate these simulations, ultimately improving the accuracy with which variations in the instrumental background can be predicted.

\begin{acknowledgments}

This work was supported by ASI (Italian Space Agency) under Contract 2019-27-HH.3-2024, and  by the ESA AREMBES Core Technology Program (CTP), contract No 4000116655/16/NL/BW. Results presented here are based, in part, upon
work funded through the European Union Seventh Framework Programme (FP7-SPACE-2013-1), under
grant agreement n. 607452, Exploring the X-ray Transient and variable Sky - EXTraS.
MB acknowledges support from the ERC CoG $\vec{B}$ELOVED, GA N.101169773.

\end{acknowledgments}

\bibliography{main}{}
\bibliographystyle{aasjournalv7}



\end{document}